# EFFICIENCY ENHANCEMENT BASED ON ALLOCATING BIZARRE PEAKS

Q. J. Hamarsheh[1], O. R. Daoud[2*], M. M. Ali[1] and A. A. Damati[3]

[1]Department of Computer Engineering,
[2]Department of Communications and Electronics Engineering,
[3]Department of Electrical Engineering,
Philadelphia University, Amman, Jordan

## ABSTRACT

*A new work has been proposed in this paper in order to overcome one of the main drawbacks that found in the Orthogonal Frequency Division Multiplex (OFDM) systems, namely Peak to Average Power Ratio (PAPR). Furthermore, this work will be compared with a previously published work that uses the neural network (NN) as a solution to remedy this deficiency.*

*The proposed work could be considered as a special averaging technique (SAT), which consists of wavelet transformation in its first stage, a globally statistical adaptive detecting algorithm as a second stage; and in the third stage it replaces the affected peaks by making use of moving average filter process. In the NN work, the learning process makes use of a previously published work that is based on three linear coding techniques.*

*In order to check the proposed work validity, a MATLAB simulation has been run and has two main variables to compare with; namely BER and CCDF curves. This is true under the same bandwidth occupancy and channel characteristics. Two types of tested data have been used; randomly generated data and a practical data that have been extracted from a funded project entitled by ECEM. From the achieved simulation results, the work that is based on SAT shows promising results in reducing the PAPR effect reached up to 80% over the work in the literature and our previously published work. This means that this work gives an extra reduction up to 15% of our previously published work. However, this achievement will be under the cost of complexity. This penalty could be optimized by imposing the NN to the SAT work in order to enhance the wireless systems performance.*

## KEYWORDS

*Orthogonal Frequency Division Multiplexing, Neural Network, Linear Codes, Wavelet, Moving Average Filter.*

## 1. INTRODUCTION

Orthogonal Frequency Division Multiplexing (OFDM) is considered as one of the main topics that is taken into consideration in order to enhance the wireless systems reliability during the last decades. This is in order to achieve a promising data rates either for the downlinks or the uplinks physical transmissions, especially after the rapid growth in using the mobile phone applications [1, 2]. Along the last decades, the researchers got from the benefits of OFDM technology and are adopted widely in the recent wireless communications technologies; especially in the wide band systems and standards, where the Mobile data traffic expected to increase by 11 times by 2018 [3, 4]. However, the Peak-to-Average Power ratio problem is considered as one of the major drawbacks that limits the efficiency of OFDM technology. The sensitivity to the nonlinear behavior especially with none constant envelops plays a vital role in affecting the powerfulness of using OFDM. Thus a spectral regrowth in adjacent channels and deformation of the signal constellation could happen. Accordingly and in order to minimize the PAPR values, many solutions were found in the literature. This will prevent the limitations in using the nonlinear devices without back-off levels, especially the power amplifiers and mixers; such as multiple signal representations, neural networks, neuro-fuzzy, selective mapping, partial transmit sequence, coding, clipping, filtering and travelling wave tube amplifiers[1,5-12].

Furthermore and to enhance the OFDM systems robustness, Multiple-Input Multiple-Output (MIMO) technology has been proposed to be combined with the OFDM techniques. MIMO technology also has a number of powerful advantages including the ability to increase the system capacity and improving the communication reliability via the diversity gain. The capacity of MIMO channels scale linearly with respect to the minimum available transmitter and receiver antennas [11, 13].

In this work, we are motivated to compare a new proposition based on a special averaging technique with the previously published work based on NN in [2]. In [2], the NN were imposed and trained based on our previously attained results in [11-13], this is in order to attain the goal of maximum amplification [14]. The NN will intelligently choose the best combination that reduces the PAPR ratio values in the transmitted OFDM signal. As a result of imposing the NN, slight enhancement has been attained at the 10 dB threshold for the BER; since it has been reduced to be $2.5\times10^{-2}$ from $1.1\times10^{-2}$ [2]. Furthermore, the MIMO technology has been imposed in order to enhance the system's capacity by introducing the use of Vertical Bell Laboratories Layered Space-Time (V-BLAST) techniques [15].

The special averaging technique will be divided into three main stages and it will make use of the test data has been collected from a funded work by the European Union (SRTD-II) and Philadelphia University; entitled by Energy Consumption Efficiency Management-Phase I (ECEM). Therefore, the wireless systems based MIMO-OFDM performance will be checked based on the bit error rate (BER) that is based on the Chernoff Union Bound and the complementary cumulative distribution function (CCDF) curves. These three stage will consist of a pre-processing stage (check the noise removal effect and to enhance the input signal characteristics), detecting the odd peaks stage (derivatives theorem, magnitude sign selection, template matching process, adaptive thresholding process), and the post processing stage (moving averaging filter).

The rest of paper is organized as follows; the introduced structure of the wireless system based on MIMO-OFDM models is defined in Section 2, the numerical and simulation results are presented in Section 3, while the last section summarizes the conclusion.

## 2. MIMO-OFDM System Description

In order to fulfill the requirements of this work, the MIMO-OFDM system's parts will cover the followings; 2/3 turbo encoder, 64 QAM, and a V-BLAST MIMO encoder. V-BLAST is used for increasing the overall throughput expressed in terms of bits/symbol, while applying the IFFT to generate the OFDM symbols [2, 15]. As a consequence for the coherent addition after the IFFT, a huge peak value may appear [1-12].

In this work, a comparison between two main proposed work will accomplish; consequently the work based on NN and will be described in subsection A, the proposed work that is based on a special averaging technique will be covered in subsection B.
In this configuration, the transmitted OFDM signal consists of a sum of 64 QAM modulated subcarriers and is shown in [1] as:

$$s(t) = \sum_{i=-\frac{N_s}{2}}^{\frac{N_s}{2}-N_s} \left[ d_{\left(i+\frac{N_s}{2}\right)} e^{(j2\pi\left(f_c - \frac{i+0.5}{T}\right))(t-t_s)} \right], \quad t_s \leq t \leq t_s + T \quad (1)$$

Here, $d_i$ can be considered as a complex input symbol resulted from the modulation stage, the carrier frequency of the $i$-th subcarrier is depicted in $f_c$, the OFDM symbol duration is defined as $T$ and the starting time of one OFDM symbol is $t=t_s$.

## 2.1. NN-Based Proposed Work

Figure 1 shows the main parts of the MIMO-OFDM transmitter stages with the addition the NN part [2].

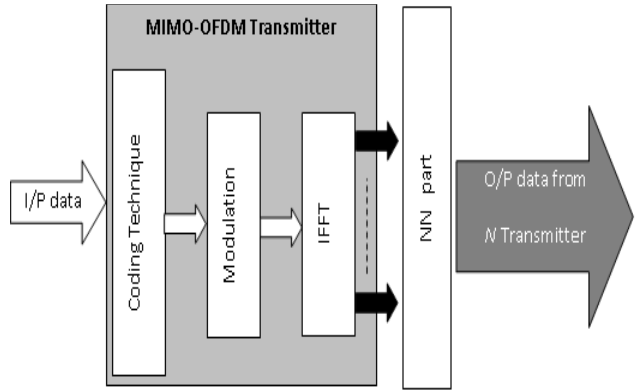

Figure 1. AN MIMO-OFDM-Based NN Transmitter

It shows that the NN part has been imposed after the IFFT stage. This is due to that the same phase addition will increase the probability of PAPR appearance [1, 2]. A high peak power equals to N times the average power occasionally appears as a result of this addition process. Thus, the PAPR definition can be written as:

$$\text{PAPR} = 10\log_{10}\left\{\frac{P_{peak}}{P_{avg}}\right\} \quad (2)$$

where, $P_{peak}$ is the maximum power of an OFDM symbol, and $P_{avg}$ is the average power.

From [11], a conclusion has been drawn that the PAPR will decrease if the average power of the OFDM symbol is decreased. This mathematical derivation was the basis of our previously published work to combat the effect of the PAPR drawback on the MIMO-OFDM systems. Imposing the NN can improve MIMO-OFDM systems performance; as it is the simplest way of linearization for RFPA. This is due to the NN ability of simultaneous BW linearization [16].

Intelligent controllers are generally self-organizing or adaptive and are naturally able to cope with the significant changes in the plant and its environment. As processes increase in complexity, they become less amenable to direct mathematical modeling based on physical laws, since they may be, distributed, stochastic, nonlinear and time-varying.

Research into intelligent systems integrates concepts and methodologies from a range of disciplines including neurophysiology, artificial intelligence, optimization and approximationtheory, control theory and mathematics. This integration of research fields has led to an emergent discipline, frequently referred to as connectionism or neuron science that inherently incorporates distributed processing concepts organized in an intelligent manner. Connectionist or neurons systems, unlike conventional techniques and self-programming, appear to be stochastic or fuzzy, heuristic and associative. An approximation to the desired mapping is constructed in intelligent or learning systems [17].

ANNs or simply NNs go by many names such as connectionist models, parallel distributed processing models, and neuromorphic systems, whatever the name, all these models attempt to achieve good performance via dense interconnection of simple computational elements. Computational elements or nodes used in neural net models are nonlinear and typically analog. The simplest node sums N weighted inputs and passes the results through a nonlinear function [2, 17, 18]. From this work, very promising results have been extracted as shown in Figure 2. The structure of the used NN can be found in Table 1 as:

Table 1. NN structure

| Functions | Description |
| --- | --- |
| Network Type | FFB-P |
| Number of layers | 3 |
| Number of Neurons | 512, 30, 512 |
| Activation Function | Bipolar-Sigmoid |
| Training Function | Error B-P |
| Performance Function & number of epochs | $10^{-3}$ and(24873, 16470) |

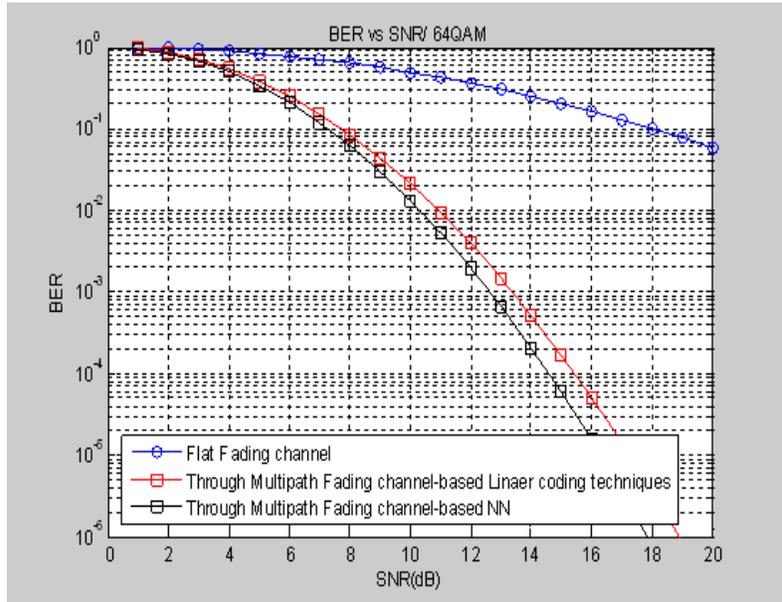

Figure 2. BER vs. SNR curves for the NN based work [2]

As mentioned previously, the previous achieved result in combatting the PAPR [11-13] has been used in the learning process of the NN part. This is in to model the MIMO-OFDM system based on multipath fading channels. The achieved results are depicted in Table 2 [2].

Table 2. Back-propagation based Powell-Beale conjugate gradient

| Learning Parameter | MSE | Number of trained OFDM symbols |
| --- | --- | --- |
| 0.1 | $6.254 \times 10^{-5}$ | 100 |
| 0.01 | $8.157 \times 10^{-5}$ | 100 |

Based on the results in [2], it can be concluded that the NN work has a slight improvement over the previously achieved results in [11-13]. From the complexity point of view, this enhancement comes in addition to the simplicity of the system structure, i.e. the NN simplifies the complexity of the work in addition to boost the performance enhancement by combatting the PAPR intelligently. In addition to the CCDF results, Figure 2 depicts a slight modification in the system's performance from the BER point of view; at 10 dB threshold as an example the BER has been reduced from $49 \times 10^{-2}$ to $12 \times 10^{-3}$. Furthermore, at 16dB threshold this enhancement has been considered and valuable since it reduces the BER from $15 \times 10^{-2}$ to $15 \times 10^{-7}$.

## 2.2. Spatial Averaging Technique-Based Proposed Work

As described in [17], Figure 3 is clearly shows the stages of the proposed work.It is based on the adaptive convolutional approach and divided into three main stages. The procedure of these stages is as follows:
- After reading the affected OFDM signal by PAPR, the pre-processing stage is started with experiences the effect of some wavelet families and their effect on clean the signal from the attached noise, such as the Biorthogonal, Daubechis, Symmlet, Coiflet and Haar wavelets. Here, three main factors have been taking into consideration in order to check the noise removal effect and to enhance the input signal characteristics, namely mean square error (MSE), signal to noise ratio (SNR) and the peak SNR (PSNR).The results from the first stage can be shown in Figure 3. It shows the results of the checked performance based on the three factors between the wavelet-based noise reduction approach and the noiseless OFDM Signals.

- The process of detecting the odd peaks (having the high PAPR) is covered in the second stage and starts with
    - Linking the idea of local maxima with the derivatives theorem and looking for the increasing-decreasing pair in the differences.
    - Convert the derived signal into a modified signal based on magnitude sign selection, where the sign indicates the trend of increasing, decreasing or remaining no change.
    - Allocating the peaks based on a template matching process making use of the discrete convolution idea. Thus, the [-1,1] pattern is used as sliding kernel window and convolved with signed data. In the output result of this process, the value of 2 appears only if the pattern matches exactly and its corresponding peak value.
    - The final step in this stage is to threshold the founded peaks by making use of a globally statistical adaptive thresholding process. Adaptively, the threshold will be determined with respect to descriptive statistics (maximum, mean and standard deviation values and constant adaptive parameter k). it is clearly found in (3) as:

$$\text{Thresh} = (m + \text{avgabs} + \text{avgdev})/k \qquad (3)$$

Here, m stands for the OFDM signal maximum value, avgabs is the mean of the absolute OFDM values, avgdev is the standard deviation, and a predefined constant k that is used in order to control the thresholding process adaptively. Thus, a thresholding limit will be varied according to the systems' limitations.

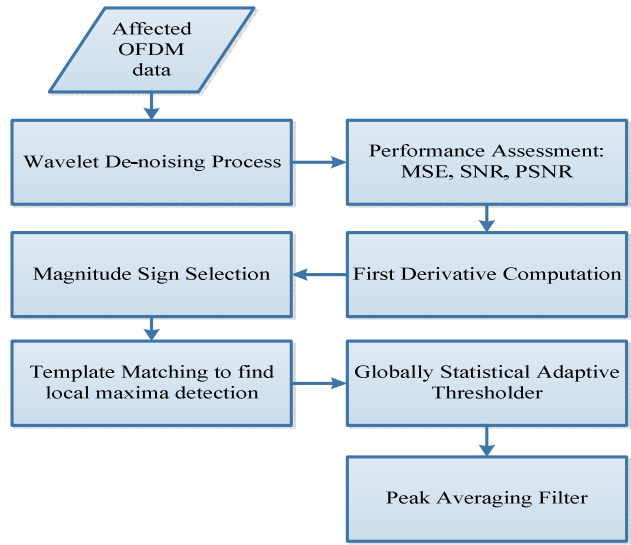

Figure 3. The flowchart of the proposed algorithm

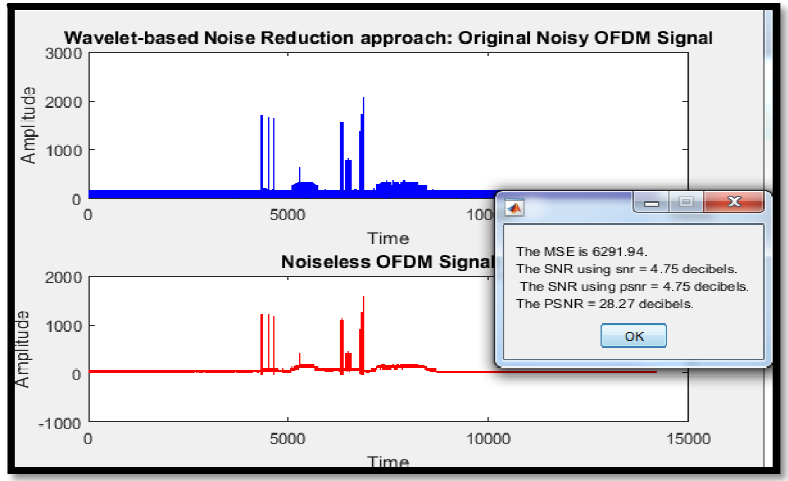

Figure 4.performance assessment methods and the wavelet-based noise reduction approach.

- The final stage deals with the post processing of the thresholded founded peaks making use of averaging filters (AF) such as the simple-AF, exponential-AF and Weighted-AF. The moving averaging filter is used in order to deal with the peak values with special treatment found in (4). These peaks will be replaced with a certain value that is extracted based on its surrounded values. The modified affected OFDM sample will be represented as:

$$OFDM_{n_{mod}} = (OFDM_{n-1} + OFDM_n + OFDM_{n+1})/3 \qquad (4)$$

Where $OFDM_{n_{mod}}$ is the modified OFDM sample, $\{...\}_{0 \leq n \leq N-1}$ is the number of OFDM sample, $OFDM_n$ is the affected OFDM sample.

Figure 5 depicts the results of the proposed procedure for detecting the large peaks and thresholding them below a predefined threshold.

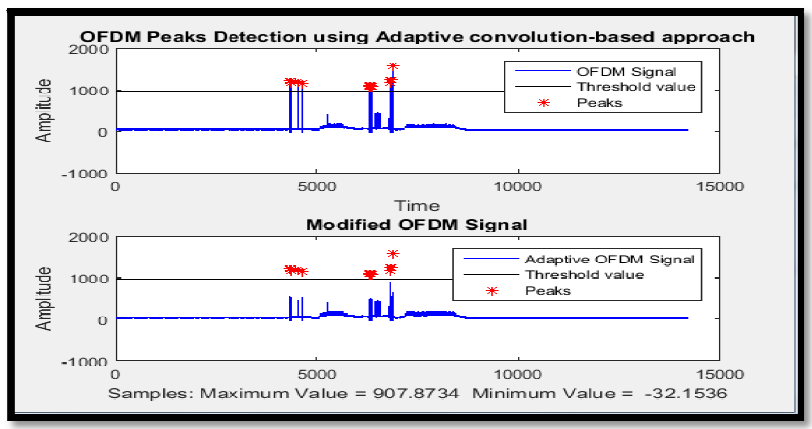

Figure 5. OFDM peaks detection using adaptive convolution-based approach

In this Figure, the red stars defining the peaks above certain threshold. These peaks are clearly treated in the modified OFDM signal part by making use of equation (4).

## 3. SIMULATION RESULTS AND DISCUSSION

Comparison between the previously published work that is based on the NN and the new work based on a special averaging technique has been made. In order to accomplish this target, a MATLAB-based simulation has been performed to imitate both of block diagram in Figure 1 and the flowchart of Figure 3. The results of this simulation will compare the OFDM performance based on the different two proposed techniques keeping in mind the two main factors; BER and CCDF. This is in addition to taking into consideration identical channel conditions with the theoretical representation. Furthermore, real signals will be extracted from the funded project by both of European Union under the SRTD-II and Philadelphia University. This is in addition to the theoretically initiated data. The system parameters were chosen to cover the following:

- 512 subcarrier
- 64-QAM modulation technique
- 2/3 inner coding rate
- 1/4 Guard interval duration
- V-BLAST MIMO coding technique

This is in addition to keep in mind the NN structure settings that shown previously in Table 1 and Table 2, which they were extracted experimentally in previous work [2].

Figures 6 and 7 show the simulation results based on CCDF curves and the BER ones, respectively. These results cover the comparison between the different two works. Furthermore and in order to check its performance, Table 3 shows a comparison that has been made with previously published work. From the achieved results, the OFDM system QoS has been checked and verified from both of two different kinds of data.

Table 3. The Simulation results of the proposed technique compared to the literature work for a 64-QAM modulation technique.

| Input Data | CCDF (2%) | | Additional Reduction over literature (%) | | |
|---|---|---|---|---|---|
| | Conventional PAPR (dB) | Proposed Work (dB) | Clipping | SLM | PTS |
| Theoretical | 20 | 8.3 | 75 | 53.3 | 20 |
| ECEM | 22 | 8 | 69 | 54 | 21 |

It is clearly shown that there is a huge improvement achieved from the proposed work; at a probability of 2% the PAPR has been reduced to around 8 dB instead of 20 dB. Moreover and comparing to the work in the literature, the attained improvement falls in the range of 20% to

International Journal of Wireless & Mobile Networks (IJWMN) Vol. 8, No. 4, August 2016

75% more enhancements over the three of them. The best values were over the clipping technique and the worst were over the PTS technique.

Figure 6 shows the CCDF comparison between the NN based work and the special averaging technique based work. These curves contain the thresholds that the probability of the PAPR will not exceed. The limitation for such simulation as mentioned earlier.

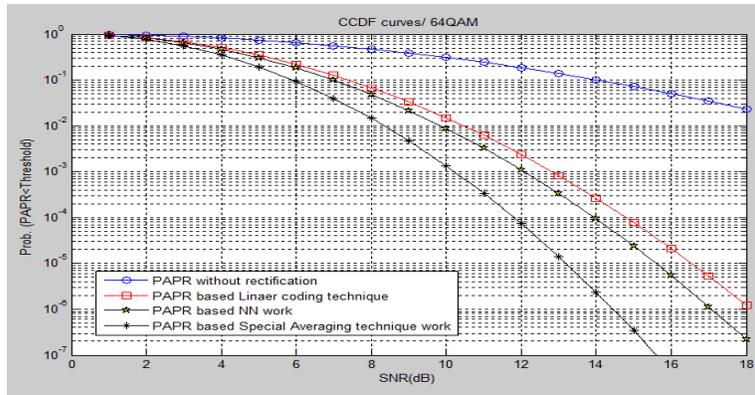

Figure 6. PAPR combating proposed work comparison

Figure 6 shows that a very promising results especially after exceeding the 12 dB threshold. It is clearly shown that before 12 dB there is an enhancement; however it cannot reflect the imposed work weight; as an example at 8 dB the probability comes between $1.5 \times 10^{-2}$ to around $6 \times 10^{-2}$. At 16 dB threshold, the probability of using the SAT gives exceeds the $10^{-7}$, while the NN based work does not exceed the probability of $7 \times 10^{-6}$ and the Linear coding one has the probability $1.8 \times 10^{-5}$.

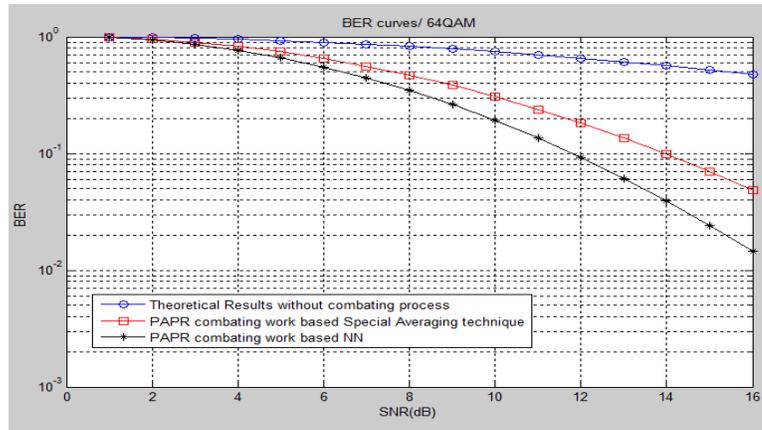

Figure 7. BER curves comparison

Figure 7, shows that nevertheless the SAT work gives a better enhancement in the CCDF curves, the NN gives better BER curves. This is due to that the NN is adaptive and compatible with the systems behavior changings. Here, the BER at 16 dB as an example is enhanced from $5.3 \times 10^{-2}$ to $1.5 \times 10^{-2}$ and fall between enhancement ratios from 18.5% to 72%.

## 3. CONCLUSIONS

In this paper, a new work has been proposed based a special averaging technique. In order to validate its efficiency, it has been compared to a previously published work that is based on neural networks, where the learning process of the proposed NN has been accomplished based on the previously extracted data from our previous published work.

From the achieved results, the OFDM system QoS has been checked and verified from both of two different kinds of data. It is clearly shown that there is a huge improvement achieved from the proposed work; at a probability of 2% the PAPR has been reduced to around 8 dB instead of 20 dB. Moreover and comparing to the work in the literature, the attained improvement falls in the range of 20% to 75% more enhancements over the three of them. The best values were over the clipping technique and the worst were over the PTS technique. Furthermore, at At 16 dB threshold, the probability of using the SAT gives exceeds the $10^{-7}$, while the NN based work does not exceed the probability of $7 \times 10^{-6}$ and the Linear coding one has the probability $1.8 \times 10^{-5}$. From the other side, the NN gives better BER curves. This is due to that the NN is adaptive and compatible with the systems behavior changings. Here, the BER at 16 dB as an example is enhanced from $5.3 \times 10^{-2}$ to $1.5 \times 10^{-2}$ and fall between enhancement ratios from 18.5% to 72%.
As a conclusion from the previous results, the proposed work have proved their reliability in overcoming the effect of PAPR. This is in addition to that the use of NN reduces the added complexity at the receiver side since there is no need to send a control data that has been sent in the previously published work.

### ACKNOWLEDGEMENTS

The authors would like to acknowledge the financial support from the SRTD-II projects under the umbrella of the European Union support and from the Philadelphia University to build the project entitled "Energy Consumption Efficiency Management (ECEM)", which is managed by the Higher Council for Science & Technology.

## Authors

Qadri J. Hamarsheh has received the master degree of Computer Machines, Systems and Networks from the department of Computer Engineering, Lviv Polytechnic Institute in 1991. He obtained his Ph.D degree from Lviv National University "LvivskaPolytechnica", Ukraina in 2001. Currently he is working as assistant professor of Computer Engineering, Philadelphia University-Jordan. He is in teaching since 2001. His areas of interest include Digital Signal Processing (DSP), Digital Image, Speech Processing, Object-Oriented Technology and Programming Languages, Internet Technology and Wireless Programming.

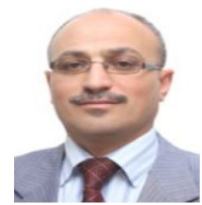

Omar R. Daoud has achieved the PhD in the field of Communication and Electronics Engineering at DMU/ UK 2006. He joined Philadelphia University in 2007 as Assistant Professor. His current work is about achieving the Quality of Service for the 4th Generation of the Wireless and Mobile Communication Systems by combining the advantages of the OFDM and the multiple antenna technology. He is the Assistant Dean in the Faculty of Engineering in addition to the Head of Communications and Electronics engineering department. Moreover, and in March 2012 he has promoted to the associate professor rank.

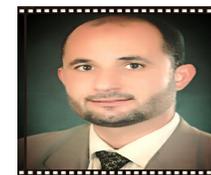

Mohammed M. Ali, received the BSc, MSc and PhD degrees in computer and control engineering from the university of Technology, Iraq in 1981, 1993, and 1998 respectively. He is currently an assistant professor in the department of computer engineering at Philadelphia University, Jordan. His research interest includes fuzzy logic, neural networks, image processing, and computer interfacing. He has 16 published papers related to real-time computer control applications. 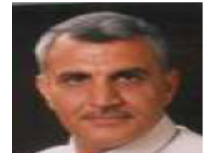

Ahlam A. Damati has achieved the MSc in the field of Electrical Engineering/ Communications from The University of Jordan/ Jordan 2002. She had worked as a lecturer in different universities since 2003; UJ and GJU, while she joined Philadelphia University in 2014 as a lecturer in the Electrical Engineering department. Her research interests are in achieving the Quality of Service for the wireless new technologies based on a variety of digital signal processing techniques. 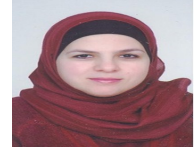